# SmartChoices: Augmenting Software with Learned Implementations

Daniel Golovin
dgg@google.com
Google DeepMind

Gábor Bartók
bartok@google.com
Google Research

Eric Chen*
eric@reka.ai
Reka AI

Emily Donahue
heyitsemily@google.com
Google DeepMind

Tzu-Kuo Huang
tzukuoh@google.com
Google DeepMind

Efi Kokiopoulou
kokiopou@google.com
Google Research

Ruoyan Qin
ruoyan@google.com
Google DeepMind

Nikhil Sarda
nikhilsarda@google.com
Google

Justin Sybrandt
jsybrandt@google.com
Google DeepMind

Vincent Tjeng
vtjeng@google.com
Google DeepMind

## ABSTRACT

In many software systems, heuristics are used to make decisions — such as cache eviction, task scheduling, and information presentation — that have a significant impact on overall system behavior. While machine learning may outperform these heuristics, replacing existing heuristics in a production system safely and reliably can be prohibitively costly. We present SmartChoices, a novel approach that reduces the cost to deploy production–ready ML solutions for contextual bandits problems. SmartChoices' interface cleanly separates problem formulation from implementation details: engineers describe their use case by defining datatypes for the context, arms, and feedback that are passed to SmartChoices APIs, while SmartChoices manages encoding & logging data and training, evaluating & deploying policies. Our implementation codifies best practices, is efficient enough for use in low–level applications, and provides valuable production features off the shelf via a shared library. Overall, SmartChoices enables non–experts to rapidly deploy production–ready ML solutions by eliminating many sources of technical debt common to ML systems. Engineers have independently used SmartChoices to improve a wide range of software including caches, batch processing workloads, and UI layouts, resulting in better latency, throughput, and click-through rates.

*Work done while at Google Research.

## 1 INTRODUCTION

Modern deep learning models power an increasing range of products and services, such as search, recommendation and discovery systems, and online advertising. However, training and deploying these models in production systems is often fraught with new failure modes and distinct forms of technical debt [27].

This paper re-envisions the workflows to deploy machine learning in large-scale production systems, with an eye towards significantly reducing engineering effort and scope for errors. The result, *SmartChoices*, uses ML models to provide *learned implementations* within an application. We tie models directly to application behavior and eliminate separate development workflows for pipeline management.[1] SmartChoices' design occupies a fertile middle ground between grand ambitions to pervasively replace traditional software with deep learning and the hard–won lessons of veteran engineers on how to build and run reliable production systems. A simple user interface, combined with a design guaranteeing low latency, reliability & safety, enables SmartChoices users to successfully deploy ML to address a diverse range of systems problems.

## 2 SCOPE AND CAPABILITIES

Machine learning can be applied towards many different problem classes. SmartChoices learns implementations for contextual bandits problems; this covers many applications (see §6). Concretely, SmartChoices addresses the following problem class. A system is faced with a sequence of decisions. At time $t$, it is provided a *context* $x_t \in \mathcal{X}$ as input, and a set of permissible outputs $A_t$ (known as *arms* in the bandit literature). $A_t$ is a subset of the universe of arms $\mathcal{A}$. The system chooses an arm $a_t \in A_t$ and receives feedback $y_t \in \mathbb{R}^k$ indicating arm quality with respect to $k \geq 1$ metrics. The goal is to provide an implementation via a *policy* $\pi : \mathcal{X} \times 2^{\mathcal{A}} \to \mathcal{A}$ to optimize the metrics[2]; SmartChoices generates policies by learning a critic model $m : \mathcal{X} \times \mathcal{A} \to \mathbb{R}^k$ and selecting an arm based on the predicted metric values for each arm in $A_t$ (e.g., via [28]).

[1]Notably, this diverges considerably from the design philosophies of ML platforms that enable setting up arbitrary ML pipelines, such as TFX [5] & Kubeflow [17].

## 2.1 Contextual Bandits

The basic formulation for contextual bandits involves a fixed small universe of arms $\mathcal{A}$ (*i.e.,* categories encoded as enum values) and a context domain $\mathcal{X} = \mathbb{R}^d$. SmartChoices supports several additional important modeling features:

- **Time-varying arm sets**: Requiring $a_t \in A_t$ when $A_t \subset \mathcal{A}$ is supported via selection masks in the policy.
- **Mixed-type contexts**: Numeric, enumeration, and string contexts are supported via automatically-generated embedding layers in the critic model for $\pi$.
- **Arm-features**: Describing each arm with mixed-type features (henceforth "arm-features") is supported via embedding layers (as with contexts), enabling generalization across arms. This is particularly valuable when $A_t$ is complex and can completely change over time (e.g., recommending an image to accompany a news story). Concretely, given an embedding $\Phi$ of objects into $\mathbb{R}^d$ and a retrieval process for selecting a reasonably sized set of candidate objects $C(x)$ for context $x$, SmartChoices will choose among $A_t = \{\Phi(c) : c \in C(x_t)\}$ and return the object associated with the selected embedding vector.

## 2.2 Multimetric optimization

There are often several metrics of interest. SmartChoices supports multimetric optimization via *scalarizing* the predicted metrics from the critic model $m$, *i.e.,* combining them in a single scalar reward via a known *scalarization* function. Decoupling scalarization from metric predictions allows us to largely reduce multiobjective optimization to the single objective case, which simplifies infrastructure and enables us to bring additional algorithmic insights to bear.

**Pareto Frontier Search**. Multimetric optimization has inevitable tradeoffs. In applications with soft constraints, users often want to understand the set of possible tradeoffs before deciding on the desired system behavior. In particular, they want to identify the policy *Pareto frontier*, defined as follows. Call a vector $y \in \mathbb{R}^k$ *achievable* if $\exists \pi . \mathbb{E}[Y|\pi(X, A)] = y$; that is, there is a policy $\pi$ with expected metrics $y$ under the distribution of contexts, arms and rewards. For maximization metrics (where higher values are desirable), the Pareto frontier is the set of achievable metric vectors $y \in \mathbb{R}^k$ such that no other achievable metric vector $y'$ exceeds it in all metrics, *i.e.,* satisfies $y'_i > y_i$ for all $i$.

SmartChoices supports setting scalarization parameters at inference time; rapidly changing between diverse scalarizations induces efficient exploration of the Pareto frontier. The choice of scalarization function affects how effectively we can discover points on the Pareto frontier. SmartChoices supports linear scalarizations and hypervolume scalarizations [31]. Linear scalarizations (*i.e.,* linear combinations of metrics) are simple and work well when the achievable metric vectors form a convex set, while hypervolume scalarizations enable discovering arbitrarily shaped frontiers.

[2]We will refer to metrics we wish to maximize as *rewards* and those we wish to minimize as *costs*.

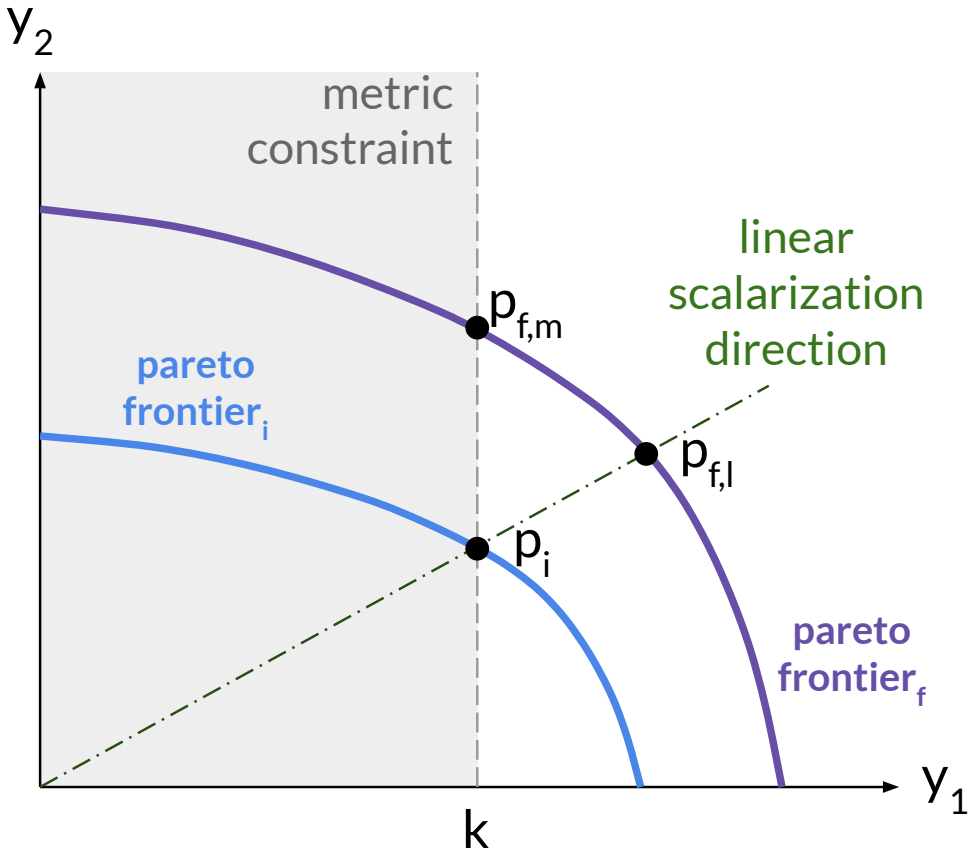

**Figure 1: Specifying the preferred tradeoff via a scalarization function and a metric–constraint can lead to the same initial behavior ($p_i$). However, if the Pareto frontier shifts (frontier$_i$ $\to$ frontier$_f$) due to changes in the operating environment (e.g., context, arm or reward distribution), the final behavior will likely be different ($p_{f,m}$ for the metric constraint $y_1 \leq k$ vs. $p_{f,l}$ for the linear scalarization direction shown). This is particularly salient if the application is sensitive to changes in a metric (e.g., if $y_1$ is latency).**

**Specifying desired system behavior**. SmartChoices supports fixing a preferred tradeoff in two ways: either by explicitly specifying the scalarization function, or specifying metric–constraints. Note that these two options behave differently when there are changes to the operating environment, as shown in Fig 1; in practice, even users with soft constraints prefer specifying metric–constraints to make system behavior more easy to reason about.

***Metric–Constraints***. SmartChoices can optimize one metric while staying within an *average budget* for the others. Formally, we can maximize reward $r : \mathcal{X} \times \mathcal{A} \to \mathbb{R}$ while keeping the average cost $c : \mathcal{X} \times \mathcal{A} \to \mathbb{R}^{k-1}$ below a target $C \in \mathbb{R}^{k-1}$, over a random stream of (context, arm set) pairs:

$$\pi^* := \arg\max_\pi \left\{ \mathbb{E}_{(X,A)} \left[ r(X, \pi(X, A)) \right] \; : \; \mathbb{E}_{(X,A)} \left[ c(X, \pi(X, A)) \right] \leq C \right\}$$

Without contexts (*i.e.,* $\mathcal{X} = \emptyset$), this is Bayesian optimization with unknown constraints [14]. With contexts, it is closely related to contextual bandits with knapsack constraints, which has known results for stochastic [3] and adversarial settings [30]. In contrast to prior work, we are interested in average budgets (e.g., serving an infinite stream of requests at bounded average latency) rather than cumulative budgets that run out (e.g., dynamically pricing a limited supply of goods for maximum revenue). Note costs, like rewards, must be *learned*; hence constraint violations during learning cannot be fully avoided. While there are ways to mitigate this, SmartChoices is not designed or intended for circumstances where individual decisions are high-stakes (e.g., selecting medical treatments).

# 3 USER INTERFACE

SmartChoices' user interface is designed to be simple and foolproof. Deploying a learned policy simply requires *(i)* specifying the problem via three types, with any additional configuration options in the single configuration file; *(ii)* adding <10 additional lines of code to the application (see Fig. 3).

## 3.1 Problem Specification

SmartChoices users specify their problem in terms of *Context*, *Arm*, and *Feedback* types. (See Fig. 2). Each type is defined as a protocol buffer (henceforth *protobuf*) message [15]; data elements within messages are described as a field with a type and a name. Users indicate field-specific modeling information via in-line annotations.

```
message ExampleContext {
  // This field is treated as a categorical, with values 0-22.
   int32 category = 1         [(opts)={num_categories: 23}];
  // String fields must specify the maximum-length string expected
  string name = 2             [(opts)={max_length: 5}];
  repeated float dims = 3     [(opts)={shape: 2}];
  // `log_only` fields are not used by critic model, but the data
  // is available for analysis (e.g., verifying the policy performs
  // well on all data slices rather than just globally)
  string debug_info = 4       [(opts)={log_only: true}];
}

message ExampleFeedback {
  // Our policy will maximize `reward` and minimize `penalty`
  float reward = 1      [(opts)={maximize_goal{}}];
  float penalty = 2     [(opts)={minimize_goal{}}];
  // `log_only` fields are not optimized over, but the data is
  // available for analysis (e.g., estimating the expected value
  // of `aux_metric` for different scalarization options)
  float aux_metric = 3 [(opts)={log_only: true}];
}
```

**Figure 2: Example context and feedback type. See comments for details on field annotations. Syntax abridged for brevity.**

Having a clear separation between the structured representations users work with (*i.e.,* protobufs) and the underlying data representation ingested by the model allows us to surface better errors (e.g., "this named field is unset") and more informative diagnostics (e.g., per-field summary statistics).

Using protobufs specifically provides several additional benefits. **Support for reflection** makes it simple to create generic components for converting any protobuf object into encoding tensors suitable for training and deploying ML models. These components eliminate the need for "glue code" and "pipeline jungles" [27] for feature extraction in user code. **Field annotations** allow users to configure parameters in-line with their field datatypes, reducing risk of "configuration debt" [27]. **Cross-language generated code** for protobuf objects enables turnkey cross-language SmartChoices support. **Compressed serialization** makes logging data memory- and bandwidth- efficient.

## 3.2 Additional Configuration Options

The three types in §3.1 are referenced in a configuration file. Users can configure additional settings for their learned policy in the same configuration file. These settings include access controls to diagnostic information, contact preferences for alerts, data retention windows, how new policies should be trained & evaluated, and so on. The single configuration file is checked into a repository; this makes it simple to compare the difference between two configurations, enforce that configuration changes undergo a full code review [27], and automatically push changes to production on a predictable cadence.

## 3.3 Using the Learned Policy

Having defined the problem, SmartChoices users instrument their code in two steps. First, users ensure that the built application has access to the compiled configuration file. In Bazel [6], this is done by creating a build rule referencing the configuration file, then declaring a dependency this build rule in the build target for the application. (The build rule also generates a test policy and basic unit tests of the problem datatypes.) Users then call the SmartChoices API in code (see example in Fig. 3); note that no separate user code is required for logging data for training.

```
auto smartchoice = CreateSmartChoice<ExampleContext, ExampleArm,
                                     ExampleFeedback>();
// protobuf with context features; see Fig. 1.
ExampleContext context = ExampleContextBuilder().SetCategory(3)
  .SetName("foo").SetDims({2, 5}).SetDebugInfo("bar");
ExampleArm selected_arm;
auto feedback_handle = smartchoice.Choose(
    // note explicit required `default_arm`.
    context, default_arm, candidate_arms, &selected_arm);
// [... user performs action with `selected_arm` ...]
// protobuf with feedback measuring impact of action; see Fig. 1.
ExampleFeedback feedback = ExampleFeedbackBuilder()
  .SetReward(3.2).SetPenalty(-0.05).SetAuxMetric(0.006);
feedback_handle.GiveFeedback(feedback);
```

**Figure 3: An abbreviated example of calling the SmartChoices API in C++.**

We highlight three key features of the API:

**A default arm is required.** This provides three advantages. First and foremost, it provides a safe fallback in case of any detected errors. Next, we can use the identity of the default arm to estimate the metrics for a policy that always chooses the default arm (henceforth the "default policy") via counterfactual policy evaluation (CPE) [7] . Finally, having the default arm allows us to implement imitation learning against the default policy and regularize to it by penalizing deviations from it. This allows us to bootstrap from a policy that achieves the current system performance.

**`selected_arm` is provided directly.** Eliminating the need for post-processing of model output is not only convenient for users but also ensures that all data required for training and analysis (e.g., $A_t$, $\pi$) is available at `Choose` and `GiveFeedback` time.

**Feedback handles are restorable.** The `feedback_handle` object required to `GiveFeedback` can be restored from an ID, allowing feedback to be provided days later and in a separate process. This is particularly useful for users who are only able to measure the quality of the decision after some delay. In addition, `Choose` and `GiveFeedback` calls can optionally be tied together by an ID provided by user code, significantly reducing the amount of infrastructure users need to add to use SmartChoices (e.g., storing a mapping from a SmartChoices ID to their ID).

# 4 IMPLEMENTATION

Many different implementations of the SmartChoices interface are possible. Our goals are to: *(i)* provide low-latency inference; *(ii)*

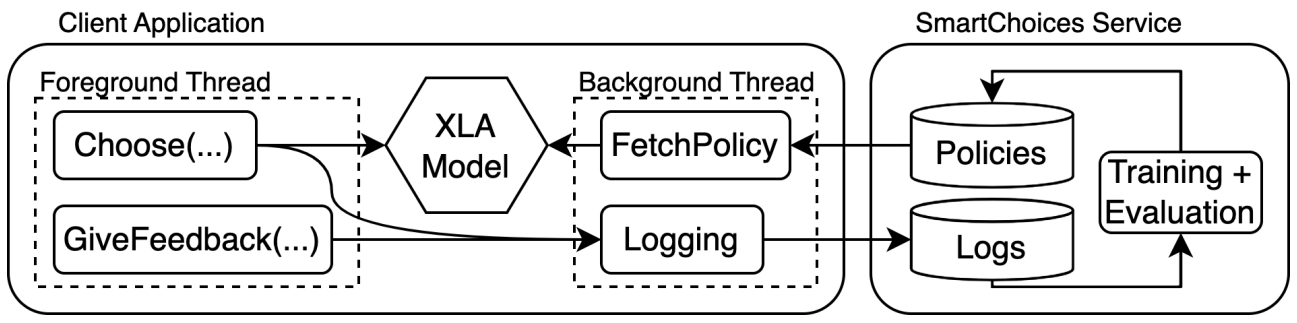

**Figure 4: Diagram of SmartChoices service implementation infrastructure. Policies are trained in a central service (see right) and sent to the client application. Inference is local to the client application and implemented using XLA.**

guarantee reliability and safety; *(iii)* eliminate the need for custom code for individual deployments.

We have two implementations: *in-process* and *service*. For in-process deployments, a background thread in client applications trains models and periodically updates the policy in use. Users choose an in-process deployment in extremely low-level settings with low latency requirements (<1 µs), where network usage is limited, or where fast policy updates (∼200 ms) are critical. This section focuses on the service implementation (see Fig. 4), in which a central multi-tenant service receives training data and provides new policies.

## 4.1 Design Principles

Our service implementation is built on three key principles.

*4.1.1 Inference is local.* SmartChoices policies are backed by local models running on CPU. Seamless switching to new policies is possible as models are instantiated by the XLA just-in-time (JIT) compiler [18]. Local inference provides latency, reliability, and safety benefits. `Choose` is non-blocking and has median latency of ∼10 µs. A trained policy is available even if the network is temporarily down. In addition, bugs causing inference errors can be identified at test time via standard unit tests employing test models, and any inference errors at runtime can be detected immediately, allowing graceful fallbacks to the user-specified default arm. Local inference *does* limit SmartChoices users to models that can fit in a single machine's RAM, but this suffices in practice to outperform existing heuristics for a wide variety of applications (see §6).

*4.1.2 All communication is asynchronous.* Background threads for communication are initialized when the `SmartChoice` object is instantiated. These periodically (i) send batches of training data by reading from a shared queue that `Choose` and `GiveFeedback` calls write to and (ii) retrieve the latest validated policy. Note: Before the policy is ready, calls to `Choose` return the default arm (§3.3).

*4.1.3 Core infrastructure is standardized.* SmartChoices' use of structured representations (§3.1) enables compatible critic models to be constructed directly from the context, arm & feedback types, with a single well-tested code path used for data preprocessing (e.g., normalizing context / arm features), model training, and policy evaluation. Data for all users is stored in the same format via use of container messages (specifically, protobuf extensions), enabling the same database and service binary to be shared by all users.

## 4.2 Service Features

Having a central service allows computationally intensive operations (e.g., policy evaluation) to be offloaded to the service. As a corollary, service users can benefit from a wide range of valuable production features without writing any additional code.

*4.2.1 Plug-and-play model training.* Getting models to train well can be complex; the service implementation abstracts much of this complexity away. First, context, arm and feedback features are scaled automatically. For each feature, we identify (from a predefined family of parameterized distributions) the distribution with the minimum statistical distance to the logged distribution. Feature values are then scaled to a target range (e.g., $[-1, 1]$) under the assumption that they were drawn from the minimal distribution. Second, hyperparameter tuning is available on demand. We integrate with an industrial-scale black box optimization platform to perform hundreds of training runs with different architectural and training configurations, identifying the parameters that minimize loss on the evaluation holdout dataset.

*4.2.2 Continuous re-training, automated evaluation & configurable validation.* SmartChoices automatically trains new policies as new training data becomes available. For each new policy, the service automatically performs a suite of evaluations using a hold-out dataset. These include computing *(i)* feature importance & prediction distributions for the critic model; *(ii)* the distribution of the selected arm; *(iii)* Pareto frontier estimates (for multimetric optimization); *(iv)* accuracy and precision plots (for boolean metrics); *(v)* comparisons to other policies (e.g., CPE) with uncertainty estimated via Poisson bootstraps [10]. Users configure validation requirements on the evaluation results to automatically determine whether a new policy is fit to serve production traffic. Evaluation and validation can be configured to run on multiple data slices, including those defined by the value of `log_only` fields.

*4.2.3 Easy access to configuration, behavior & performance information.* An automatically generated web frontend shows: *(i)* the problem specification (§3.1) and any additional settings (§3.2); *(ii)* counts of `Choose` and `Feedback` calls for each policy over time; *(iv)* logged distributions for each context, arm, and feedback field for each policy over time; *(iii)* training progress (e.g., current epoch & loss); *(v)* evaluation results & validation status on selected data slices for each policy.

*4.2.4 Automated alerts.* Users receive alerts for potential issues (e.g., logged data missing a required field, no new policy passing validation requirements for a protracted period of time, numerical errors during training, or `Choose` calls falling back to the default arm due to inference errors at runtime).

*4.2.5 Desired tradeoffs for multimetric optimization are specified as a configuration setting.* The service generates policies based on a "tradeoff configuration" specified in the user configuration file. A different tradeoff can be specified for each data slice, with a policy generated per slice specified. As discussed in §2.2, users can either specify the scalarization function directly or specify metric–constraints. Constraints can even be specified with respect to the behavior of a baseline policy (e.g., use up to 95% of the compute of the default policy). For metric–constraints, the service identifies the scalarization with the highest expected reward that also respects the constraints by searching across the range of scalarization function parameters. In both cases, having tradeoffs as a configuration setting rather than requiring users to scalarize multiple metrics into a single reward up front when providing `Feedback` accelerates experimentation. Users specifying metric–constraints further benefit from robustness to shifts in the operating environment (see Fig 1).

*4.2.6 Programmatically updated human-readable policy "tags" for managing policy rollouts.* Tags are mutable references to a policy that users can refer to directly in code with two main use-cases.

First, tags enable SmartChoices to be used within the framework of users' existing integration testing, canary, release and rollback processes. While many service users immediately deploy the most recent trained policy that passed validation ("`live`"), others choose to use environment-dependent policy tags, allowing policy updates to be first deployed in a staging environment before being used for all traffic. In an emergency, tags can also be rolled back to their previous state with a single command.

Tags also simplify experimentation & deployment for multimetric optimization. A tag is generated for each configured tradeoff and data slice (e.g., "`live_tradeoff1_sliceA`") and the tagged policy is updated each time a new model is trained. Users who need to decide their tradeoff configuration based on metrics (e.g., Daily Active Users) that are not attributable to a single choice (e.g., selecting whether to send a notification) can do so easily by running experiments referencing different tradeoff tags. Similarly, selecting different tradeoffs for different traffic slices (e.g., based on geography) is as simple as pointing to the correct tag.

# 5 EVALUATION

## 5.1 Production Readiness

Our service implementation mitigates many of the novel types of technical debt and production risks ML can create (see §1) by design. To quantify this, we evaluate our ML production readiness in Table 1 via the rubric of Breck et al. [8].

## 5.2 Safety

Even with rigorous software engineering practice, bugs will inevitably arise — via logical errors, numerical instabilities, or low level compiler bugs. Conducting model inference in-memory on CPU allows us to detect inference errors (e.g., malformed inputs, infinite predicted rewards, or errors in the XLA compiler) without overhead. If a failure is detected, we safely fall back to outputting the default arm (§4.1.1) and log data that triggers an alert (§4.2.4).

## 5.3 Fairness

As computational systems take on increasing influence in society, it has become increasingly important to understand the implications, and, ideally, design systems that encourage healthy outcomes for businesses and society at large. This presents a vast research frontier (see e.g., Chouldechova and Roth [11]) that is currently actively being explored, even at the foundational level of appropriate formal definitions of fairness[3]. Still, whatever best practices around ML

[3]Some definitions are incompatible, with known impossibility results revealing fundamental tradeoffs.

| **Data Tests** | |
|---|---|
| Feature expectations are captured in a schema. | ✓ §3.1 |
| All features are beneficial. | ✎ §4.2.3[a] |
| No feature's cost is too much. | 🙋 |
| Features adhere to meta-level requirements. | ✓ §3.3[b] |
| The data pipeline has appropriate privacy controls. | ✓ §3.2[c] |
| New features can be added quickly. | ✓ §3.1 |
| All input feature code is tested. | 🙋 |
| **Model Tests** | |
| Model specs are reviewed and submitted. | ✓ §3.2 |
| Offline and online metrics correlate. | ✓ §4.2.2[d] |
| All hyperparameters have been tuned. | ✓ §4.2.1 |
| The impact of model staleness is known. | ✎ §4.2.2[e] |
| A simpler model is not better. | ✎ §4.2.1[f] |
| Model quality is sufficient on important data slices. | ✓ §4.2.2 |
| The model is tested for considerations of inclusion. | ✎ |
| **Infrastructure Tests** | |
| Training is reproducible. | ✓[g] |
| Model specs are unit tested. | ✓ §3.3 |
| The ML pipeline is integration tested. | ✓[g] |
| Model quality is validated before serving. | ✓ §4.2.2 |
| The model is debuggable. | ✓ §4.2.3[h] |
| Models are canaried before serving. | ✓🔧 §4.2.6 |
| Serving models can be rolled back. | ✓ §4.2.6 |
| **Monitoring Tests** | |
| Dependency changes result in notification. | 🙋 |
| Data invariants hold for inputs. | ✓ §4.2.4 |
| Training and serving are not skewed. | 🙋🛡[i] |
| Models are not too stale. | ✓🔧 §4.2.4 |
| Models are numerically stable. | ✓ §4.2.4 |
| Computing performance has not regressed. | ✎ |
| Prediction quality has not regressed. | ✓ §4.2.2 |

✓ met by design (🔧 indicates configurable behavior); see referenced section for details

✎ users must handle manually based on analysis automatically performed and displayed on our frontend (§4.2.3)

🙋 user responsibility (🛡 indicates using SmartChoices simplifies passing this test)

[a] Field sensitivity charts support identification of non-beneficial features for later removal.
[b] Meta-level requirements are enforceable at the `Choose` callsite; we guarantee that no additional features will be used for training.
[c] Diagnostic information for each deployment is access controlled, with controls specified in a configuration file that is reviewed.
[d] CPE correlates with online metrics. Users can compare offline and online metrics easily in the web frontend.
[e] CPE against logged data reveals the impact of model staleness.
[f] Hyperparameter tuning on architectural parameters can surface whether simpler models perform better.
[g] These rubric tests are met by testing of our implementation.
[h] Evaluations such as feature importance, prediction distributions, distributions of selected arms, and accuracy and precision plots aid in model debugging.
[i] SmartChoices reduces the likelihood of training/serving skew by *(i)* discouraging the use of different code paths in training and inference, and *(ii)* facilitating detecting changes in the front-end (§4.2.3).

**Table 1: Measuring SmartChoices against the ML Test Score Rubric of Breck et al. [8]. SmartChoices deployments meet most of these tests by design; the relevant sections are referenced with an additional description where necessary. Tests that remain user responsibility all involve feature generation.**

fairness emerge over time, treating ML models as code and fundamentally tying models to the decisions they result in (§3) has the advantage of providing a centralized surface to design, implement, and monitor compliance with the desired behavioral constraints in production.

## 6 CASE STUDIES

A wide range of problems fit into the contextual bandits framework (§2). With the context, arms, and feedback information in place, SmartChoices users have successfully deployed a learned policy with just days of engineering time. The low latency of SmartChoices deployments enables ML to be applied towards low-level problems, and its guaranteed safety and reliability gives engineers the confidence to deploy ML in mission-critical settings. As evidence, we share a representative sample of deployments.

***Learned Cache Eviction.*** SmartChoices reduced the fraction of user-requested bytes missed in a large-scale video Content Delivery Network (CDN) cache by 9.1% at peak traffic by improving its eviction policy. The low-latency inference and continuous training provided by SmartChoices was crucial in enabling this deployment. Details on the learned eviction policy and a comparison to existing heuristic and learned cache algorithms are available at [29].

***Optimizing Compilation.*** SmartChoices enabled the XLA compiler to better optimize tile size selection [24] for high-level operations (HLO) by reducing the time required to find a good tile size. Instead of evaluating all possible tile sizes for each HLO, we evaluate a subset of candidates (1% of the total) identified as promising by SmartChoices. Across HLOs, the best tile size in the SmartChoices-identified subset provides a speedup that is is 90 – 95% of the best possible. Reducing search time per HLO enabled tile size selection to be run more frequently in response to changes in other parts of the compiler. More detail on context features and a discussion of the differences between this optimization approach and existing search-based techniques (e.g., TVM) are available at [23].

***Optimizing Thread Counts.*** SmartChoices reduced tail latency for end-user queries on a flight booking search service by optimizing thread count for parallelizable stages (e.g., processing all relevant sequences of flights). SmartChoices dynamically selected the thread count per query based on context features (e.g., number of flight sub-sequences and the source and destination regions). For each query, latency and CPU usage is provided as feedback metrics, with SmartChoices minimizing a weighted combination (§2.2) of the two. At launch, average latency reduced by 25% and 99$^{\text{th}}$ percentile latency reduced by 16%, without a significant increase in CPU cost.

***Optimizing Work Partitioning.*** SmartChoices improved data freshness for a monitoring service via dynamic work rebalancing between tasks [16]. Each task retrieves & summarizes telemetry for a list of workloads. Overly large tasks are terminated when they reach an execution deadline $t_d$, resulting in stale data for any unprocessed workloads. Conversely, overly small tasks incur unnecessary overhead. For each task, SmartChoices chose whether to `shard` the list of workloads (with 0 reward) or `execute` (with reward $t_d - t_e$, where $t_e$ is actual execution time); the reward formulation encourages sharding only large tasks. Context features include workload count, day of week, and metadata on retrieved data sources. Using SmartChoices reduced the fraction of tasks hitting $t_d$ by 53% and significantly reduced alerts on stale data, translating directly into time savings for the service's owners.

***Optimizing Refresh Rates.*** SmartChoices learned to balance resource usage with revenue for a large-scale advertisement system. In this setting, ads are cached on end-user devices, with periodic requests for "refreshed" ads made to an ad server. The server runs an ad auction and returns an ad for only a portion of requests; the remaining requests are dropped and the end-user device continues to display the cached ad. Computational resources used to run an auction are considered wasted if the end–user does not view the refreshed ad (i.e., a new auction is run before the next ad view). SmartChoices was used to choose whether to run the ad auction for each request. Context features include high-level information about the request, with two binary metrics provided as feedback: whether we ran the ad auction, and whether the ad returned was viewed (identically 0 if the ad auction was not run).

SmartChoices was configured using metric constraints to minimize ad auctions while maintaining the average number of ad views per request above a threshold (§2.2). Business goals necessitated a different threshold for different platforms (e.g. Android, iOS, and desktop); we easily expressed this in the configuration as different tradeoffs for each data slice (§4.2).

The ease of adjusting these thresholds allowed our users to deploy SmartChoices quickly and with confidence. First, the number of auctions was reduced by 12% without any changes to ad revenue. Next, the frequency of ad refresh requests from end-user devices was increased, with partially–offsetting adjustments made to the metric–constraint on average ad views per request. This increased ad views by 4.8% at the cost of a 2.2% increase in auctions. SmartChoices also adapted easily to the removal of a separate upstream heuristic that filtered some requests, increasing ad views by 1.4% at the cost of a 4% increase in auctions.

***Optimizing User Interfaces.*** SmartChoices improved user engagement metrics for a variety of applications by identifying the best of a set of human-designed options for the user interface. Example aspects of the interface optimized include notification strings and text displayed during mobile device onboarding. Engagement metrics improved by 2 – 10%, but SmartChoices' key improvement over the status quo of A/B testing was a significant reduction in the engineering time required to deploy and maintain optimizations. SmartChoices' simple user interface (§3) and plug-and-play model training (§4.2.1) reduced the effort required to optimize each aspect of an interface. Support for multiple context features (§2.1) eliminated the separate A/B experiments that had been used to optimize context features one by one. Finally, continuous retraining (§4.2.2) enabled automatic adaptation to external changes without periodic new A/B experiments.

## 7 RELATED WORK

Beyond the examples discussed in §6, contextual bandits have been applied across industry to a wide range of problems, including personalizing products [4], improving recommendations [12, 19, 26], and responding well in ambiguous contexts [21, 25]. The wide range of applications speaks to the huge potential impact of an approach like SmartChoices that accelerates improving systems with ML.

| | DS | R | L | PSS | SC | |
|---|---|---|---|---|---|---|
| *Multimetric Optimization* | | | ✓ | | ✓ | (§2.2) |
| *Built-in Fallback Logic* | | | | | ✓ | (§3.3) |
| *Local Inference* | ✓ | | | ✓ | ✓ | (§4.1) |
| *Training on Data from Multiple Machines* | ✓ | ✓ | ✓ | | ✓ | (§4.2) |
| *Policy Rollouts can be Gated on CPE* | ✓ | ✓ | ✓ | | ✓ | (§4.2.2) |
| *Median Inference Latency (reported) / µs* | 200† | N/A | 2000‡ | 0.004 | 10 | (§4.1) |

**Table 2: A comparison of the key features of SmartChoices & other projects deploying ML for decision making within production systems. SmartChoices has the lowest decision latency of the service-based systems, making it best suited for using data from an entire cluster for low-level system optimizations. †: average latency; ‡: excludes feature extraction latency of** 45 ms. ***DS: Decision Service, R: ReAgent, L: Looper, PSS: Prediction System Service, SC: SmartChoices.***

Two classes of problems similar to contextual bandits are Bayesian optimization and reinforcement learning (RL). Bayesian optimization focuses on the low-data regime where gathering feedback is expensive; in contrast, we focus on settings with more abundant data. In the RL setting, the arm selected at time $t$ affects the next context $x_{t+1}$. Since a large class of practical optimizations can be framed as a contextual bandit problem, we have not yet needed to support RL, though it would be straightforward to add.

Carbune et al. [9] presented an earlier prototype of SmartChoices. The version we present here has some significant differences based on our production experience, most notably a focus on contextual bandits over general RL, assorted safety features, and a service architecture. Associated semantics for SmartChoices were analyzed in Abadi and Plotkin [1].

Several other projects have explored how to effectively deploy ML for decision making within production systems. We summarize key differences in Table 2, with additional details below.

**Microsoft's Decision Service** [2] shares many design principles & goals with SmartChoices. SmartChoices' lower decision latency makes it better suited for low-level system optimizations. **ReAgent** [13] (prev. Horizon) is a platform for deploying RL in production, with similar goals to SmartChoices. SmartChoices provides greater ease of use (e.g., providing more automation for data preprocessing, training, and model updates). **Looper** [20] is an end-to-end platform for training and deploying machine learning models, with emphasis on optimizing over "product goals": metrics that summarize the aggregate effect of many decisions. Looper uses both per-decision metrics and product goals to optimize over a class of parameterized "strategy blueprints" using a sequence of A/B tests selected via Bayesian optimization. Unlike Looper, SmartChoices does not have a direct notion of product goals and is not tightly integrated with an A/B experiment framework. However, optimizing per-decision metrics that are a proxy for the product goal (combined with a grid search over scalarization parameters in the multimetric case) has typically sufficed for SmartChoices users. Last, the **Prediction System Service** [32] focuses on low-latency predictions for systems applications. It is similar to our in-process implementation (§4), with logging and training occurring on a single machine. However, in contrast to SmartChoices, it focuses on predictions, without an explicit notion of data gathering or exploration. Using service SmartChoices also enables training on data from an entire cluster and supports the features in §4.2.

Finally, Natarajan et al. [22] present **Programming by Rewards** (PBR), which synthesizes decision functions as if-then-else programs that are checked in. This has advantages in terms of interpretability, speed, and avoiding additional dependencies; however, there are no affordances for adapting policies to changing environments over time. In addition, the class of policies supported are restricted to either filling in values from a user-provided template (*i.e.,* programs with missing constants), or programs representable by fixed–depth decision trees.

## ACKNOWLEDGMENTS

This work benefited from the wisdom of many colleagues engaged in the development of machine learning in mature production systems. We especially wish to thank Jeff Dean, Sanjay Ghemawat, Jay Yagnik, Andrew Bunner, George Baggott, Jesse Berent, Jimmy Voss, Changfeng Liu, Ben Solnik, Alex Grubb, Weikang Zhou, Eugene Kirpichov, Arkady Epshteyn, Ketan Mandke, Wei Huang, and Eugene Brevdo for thoughtful input into the design and goals of the SmartChoices project. Lastly, we would like to thank our many colleagues who integrated SmartChoices into their projects and championed this effort within their respective teams.